\documentclass[showpacs,twocolumn]{revtex4}
\usepackage{graphicx}

\newcommand{\diff}{\mathrm d}

\renewcommand{\v}[1]{{\bf #1}}

\begin{document}
\title{
Ergodicity of Thermostat Family of Nos\'e--Hoover type
}

\author{Hiroshi Watanabe,$^{1}$\footnote{E-mail: hwatanabe@is.nagoya-u.ac.jp}
Hiroto Kobayashi$^2$}

\affiliation{
$^1$Department of Complex
Systems Science, Graduate School of Information Science,
Nagoya University, Furouchou, Chikusa-ku, Nagoya 464-8601, Japan}

\affiliation{
$^2$ Department of Natural Science and Mathematics,
Chubu University, Kasugai 487-8501, Japan
}

\begin{abstract}
One-variable thermostats are studied 
as a generalization of the Nos\'e--Hoover method
which is aimed to achieve Gibbs' canonical distribution with conserving
the time-reversibility.
A condition for equations of motion
for the system with the thermostats is derived in the form of a partial differential equation.
Solutions of this equation construct a family of thermostats
including the Nos\'e--Hoover method as the minimal solution.
It is shown that the one-variable thermostat coupled with the one-dimensional harmonic oscillator
loses its ergodicity with large enough relaxation time.
The present result suggests that multi-variable thermostats are required
to assure the ergodicity and to work as heatbath.
\end{abstract}


\maketitle


One of the important issues of recent simulation studies
is achieving the canonical distribution for the system 
at the desired temperature.
Traditional molecular dynamics (MD) simulation has been performed
on the basis of the Hamiltonian form which gives the microcanonical distribution.
In the microcanonical system, it is difficult to control the temperature
since all we can do is to set up the initial configuration.
Therefore, we need canonical MD which is defined as a method to achieve 
the canonical distribution for the system.
In addition to the above, some properties are also desired;
(i) Autonomous dynamics, {\it i.e.}, the equations of motion should be
a closed form and the dynamics should be deterministic;
(ii) Time-reversibility;
(iii) Ergodicity.

Many methods are proposed to control temperature in MD simulations.
The first method controlling temperature was proposed by Woodcock~\cite{Woodcock}.
While this method is very simple, it is non-autonomous since it involves
artificial velocity-scaling.
The autonomous method is proposed on the basis of
the variational principle with constraint
which is refered to the Gaussian thermostat~\cite{GaussianThermostat}.
This thermostat gives the canonical distribution for potential energy
with conserving kinetic energy.
Nos\'e proposed the extended system method which gives the canonical
distribution for the total energy~\cite{Nose}.
This method was reformulated to a simple form by Hoover,
and it is now refered to the Nos\'e--Hoover method~\cite{NoseHoover}.
The Nos\'e--Hoover method achieves the canonical distribution for a given system
by adding one degree of freedom~\cite{NoseReview}.
The Hamiltonian formulations have been also proposed~\cite{Dettmann,NosePoincare}.
Recently, Hoover {\it et al.} showed that the deterministic thermostats 
can be applied for far-from-equilibrium problems~\cite{HooverAoki}
and Kusnezov {\it et al.} extended the Nos\'e--Hoover dynamics
to classical Lie algebras~\cite{Kusnezov2}.

While the Nos\'e--Hoover method is convenient to study various isothermal systems,
it is found that the method sometimes loses its ergodicity, and consequently fails to achieve
the canonical distribution.
In order to improve the ergodicity of the Nos\'e--Hoover method,
some extended methods are proposed~\cite{NoseHooverChain,KineticMoments, Kusnezov1}.
In Ref.~\cite{Kusnezov1}, 
Kusnezov {\it et al.} proposed the general formulation of the extended Nos\'e--Hoover method and
concluded that two additional degrees of freedom are enough to make a system ergodic.
However, we have not had the reason why the multi-variable thermostat achieves 
the ergodicity while the single-variable ones lose~\cite{Branka}.
Therefore, we study the ergodicity of general single-variable thermostats
in order to investigate when and why the system loses its ergodicity.
In the present Letter,
we first derive the condition which the equations of motion should satisfy
to achieve Gibbs' canonical distribution
and we give the general expressions for the one-variable thermostats
extended from the Nos\'e--Hoover method.
Then we show that the one-variable thermostat coupled with the
one-dimensional harmonic oscillator loses its ergodicity for large relaxation time.


Consider the distribution function $f$ and the state vector $\v{\Gamma}$
of a phase space. Let $H(\v{\Gamma})$ be a pseudo Hamiltonian
describing the energy of the system at $\v{\Gamma}$.
The distribution function is normalized as
\begin{equation}
\int f \diff \v{\Gamma} = 1, \label{eq_normalization}
\end{equation}
and the internal energy of the system is given by $U$ as
\begin{equation}
\int H f \diff \v{\Gamma} = U. \label{eq_internal_energy}
\end{equation}
The entropy of the system is defined by
$$
S = - k_{\rm B} \int f \log f \diff \v{\Gamma}
$$
with the Boltzmann constant $k_{\rm B}$.
The equilibrium state is obtained by maximizing the entropy 
under the conditions Eqs.~(\ref{eq_normalization}) and (\ref{eq_internal_energy}),
and the canonical distribution is obtained to be
\begin{equation}
f = Z^{-1} \exp{(-\beta H)} \label{eq_canonical_distribution}
\end{equation}
with the partition function $Z \equiv \int \exp{(-\beta H)} \diff \v {\Gamma}$.
In order to perform an MD simulation, equations of motion for $\v{\Gamma}$ must be explicitly given.
The equation of continuum for $f$ and $\dot{\v{\Gamma}}$ is
$$
\frac{\partial}{\partial \v{\Gamma}} ( \dot{\v{\Gamma}}  f ) = 0,
$$
where $\partial/\partial \v{\Gamma}$ denotes the divergence and the distribution is assumed to be stationary.
In order to achieve the canonical distribution Eq.~(\ref{eq_canonical_distribution}),
we have the following condition for $\dot{\v{\Gamma}}$ as
\begin{equation}
\frac{\partial \dot{\v{\Gamma}}}{\partial \v{\Gamma}} = \beta \dot{H}
= \beta \frac{\partial H}{\partial \v{\Gamma}} \dot{\v{\Gamma}}. \label{eq_divergence}
\end{equation}
Note that the flow of this dynamics is compressible since the divergence of $\dot{\v{\Gamma}}$ is not zero,
while the flow is incompressible in the microcanonical system~\cite{Kusnezov1}.


The equations of motion satisfying Eq.~(\ref{eq_divergence}) achieve the canonical
distribution for arbitrary chosen $H$, provided that the system is ergodic.
In most cases, the system of interest is described by a Hamiltonian.
Let $H_0$ be such Hamiltonian defined in a $2N$-dimensional phase space
$\v{\Gamma}_0 = (q_1,\cdots,q_N,p_1,\cdots,p_N)$
which is a subspace of $\v{\Gamma}$,
that is,
$
\v{\Gamma} =  \v{\Gamma}_0 \otimes \v{\Gamma}_{\perp}.
$
The distribution function of the subsystem is obtained by 
the projection from $\v{\Gamma}$ onto $\v{\Gamma}_0$ as
\begin{equation}
f_0(\v{\Gamma}_0) = \int f \diff \v{\Gamma}_{\perp}. \label{eq_projection}
\end{equation}
If the pseudo Hamiltonian $H$ is chosen as
$$
H(\v{\Gamma}) =  H_0(\v{\Gamma}_0)+ H_\perp(\v{\Gamma}_{\perp}), 
$$
then the distribution function becomes
\begin{equation}
f(\v{\Gamma}) =  f_0(\v{\Gamma}_0) f_\perp(\v{\Gamma}_{\perp}), \label{eq_embedded}
\end{equation}
since $f \propto \exp{(-\beta H)}$.
With Eqs.~(\ref{eq_projection}) and (\ref{eq_embedded}),
we obtain the canonical distribution for the given Hamiltonian to be
$$
f_0 = Z_0^{-1} \exp{(-\beta H_0)} 
$$
with $ Z_0^{-1} \equiv Z^{-1} \int \exp{(-\beta H_{\perp})} \diff \v{\Gamma}_{\perp}$.

\begin{figure}[tbph]
\includegraphics[width=8cm]{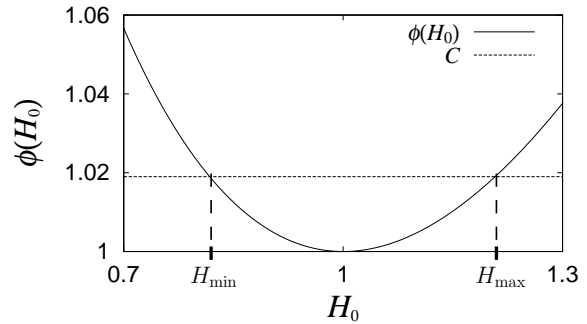}
\caption{
The range of the energy $H_0$.
The function $\phi(H_0) = H_0 - (m+1)\beta^{-1} \log H_0$ and $C$ are plotted
for $m=0$, $\beta = 1$, and $C = 1.019$.
The minimum and the maximum values are determined as solutions of 
$\phi(H_0) = C$. Note that this equation always has
two positive solutions $H_{\rm min}$ and $H_{\rm max}$.
From the inequality~(\ref{eq_inequality}),
we can estimate the range of the energy as $0.815 \leq H_0 \leq 1.211$.
}
\label{fig_inequality}
\end{figure}

Even if the pseudo Hamiltonian $H$ is explicitly given,
there are various choices of the dynamics.
In the present Letter, we consider one-variable thermostats
as extension of the Nos\'e--Hoover method
since it is favorable to simulate systems with less degrees of freedom.
Then the total phase space is defined by $\v{\Gamma} = (q_1,\cdots,q_N,p_1,\cdots,p_N) \otimes (\zeta)$
with the additional degree of freedom $\zeta$.
In order that the distribution function of the subsystem exists,
the integration of the total distribution function over $\zeta$ should converge as
$$
\int_{-\infty}^{\infty} \exp{(-\beta H_{\perp})}  \diff \zeta  < \infty. 
$$
The simplest function satisfying the above condition is $H_{\perp}(\zeta) = \zeta^2/2$,
and therefore we consider the pseudo Hamiltonian
\begin{equation}
H = H_0 + \frac{1}{2} \tau^2\zeta^2 \label{eq_pseudo_h}
\end{equation}
with the relaxation time $\tau$.
For the simplicity, we consider the subsystem $H_0$ with one degree of freedom
hereafter. 
The following arguments are not changed in the case with many degrees of freedom.

Consider the following equations of motion:
\begin{eqnarray}
\dot{p} &=& -\displaystyle\frac{\partial H_0}{\partial q} - g, \label{eq_nose_hoover_ex1} \\ 
\dot{q} &=& \displaystyle\frac{\partial H_0}{\partial p}, \label{eq_nose_hoover_ex2}\\
\dot{\zeta} &=& F[g], \label{eq_nose_hoover_ex3}
\end{eqnarray}
which are simple extension of the Nos\'e--Hoover method.
The function $g(p,\zeta)$ is a friction term which is $p\zeta$ in the Nos\'e--Hoover method.
The time derivative of $\zeta$ depends on $g$
and it is determined from the condition Eq.~(\ref{eq_divergence}).
From Eqs.~(\ref{eq_divergence}) and (\ref{eq_pseudo_h}), we have the following
partial differential equation:
\begin{equation}
\tau^2
\left(
\beta \zeta - \frac{\partial}{\partial \zeta}
\right) \dot{\zeta}
= 
\left(
\beta p - \frac{\partial}{\partial p}
\right) g ,
\label{eq_condition}
\end{equation}
which $\dot{\zeta}$ should satisfy.
Here we assumed the natural Hamiltonian form $H_0 = p^2/2 + V(q)$ with the potential energy $V$.
The function $g$ depends only on $p$ and $\zeta$ since Eq.~(\ref{eq_condition}) does not contain $q$.
The solution of the equation gives $\dot{\zeta}$ as a function of $p$ and $\zeta$,
and then the equations of motion are closed and become autonomous.
The solution of Eq.~(\ref{eq_condition}) for the case
$\partial \dot{\zeta}/\partial \zeta = 0$ is given in Ref.~\cite{Kusnezov1}.
In the present Letter, we study more general solutions both for $\partial \dot{\zeta}/\partial \zeta \neq 0$.

The equations of motion of the Nos\'e--Hoover method are
time-reversible with the operation $p \rightarrow -p$, $q \rightarrow q$,
and $\zeta \rightarrow -\zeta$.
Similarly, the equations of motion Eqs.~(\ref{eq_nose_hoover_ex1})--(\ref{eq_nose_hoover_ex3})
are also time-reversible when $g \rightarrow g$.
Therefore, the function $g$ is a linear combination of $p^k \zeta^l (k \ge 0, l \ge 0, k+l = 2, 4, 6, \cdots)$.
Here we assume that $g$ does not contain the negative power of $p$ and $\zeta$
for the stability of the dynamics.
In the case that both $k$ and $l$ are even, it is difficult to control temperature since
$p^k \zeta^l$ becomes positive semi-definite \cite{even}.
Therefore, we consider only odd cases as $g = p^{2m+1} \zeta^{2n+1} ~(m,n = 0, 1, 2, \cdots)$,
and then we obtain the expression for $\dot{\zeta}$ as
\begin{equation}
\dot{\zeta} 
=  \frac{1}{\tau^2} z_n \left( p^{2m+2} - \frac{2m+1}{\beta} p^{2m} \right), \label{eq_dotzeta}
\end{equation}
where the function $z_n(\zeta)$ is the solution of the following 
ordinary differential equation:
$$
\left( \beta \zeta  - \frac{\diff}{\diff \zeta} \right) z_n = \beta \zeta^{2n+1}, 
$$
and it is explicitly expressed as
$$
z_n = \left( \frac{2}{\beta} \right)^n n! \sum_{k=0}^{n} \frac{1}{k!} \left( \frac{\beta \zeta^2}{2} \right)^k. 
$$

Equation~(\ref{eq_dotzeta}) gives the Nos\'e--Hoover method as the minimal solution
with $(m,n) = (0,0)$.
The case $(m,n)=(1,0)$ gives the thermostat
controlling only the second moment of the kinetic energy $\left< K^2 \right>$ as
\begin{eqnarray*}
g &=& p^3 \zeta,\\
\dot{\zeta} &=& \displaystyle \frac{1}{\tau^2} p^2 \left( p^2 - \frac{3}{\beta} \right).  
\end{eqnarray*}
The kinetic--moments method~\cite{KineticMoments} is obtained from $g = g_1(p,\zeta) + g_2(p,\xi)$
with $g_1$ for $(m, n)=(0, 0)$ and $g_2$ for $(m, n)=(1, 0)$.


\begin{figure}[tb]
\includegraphics[width=7cm]{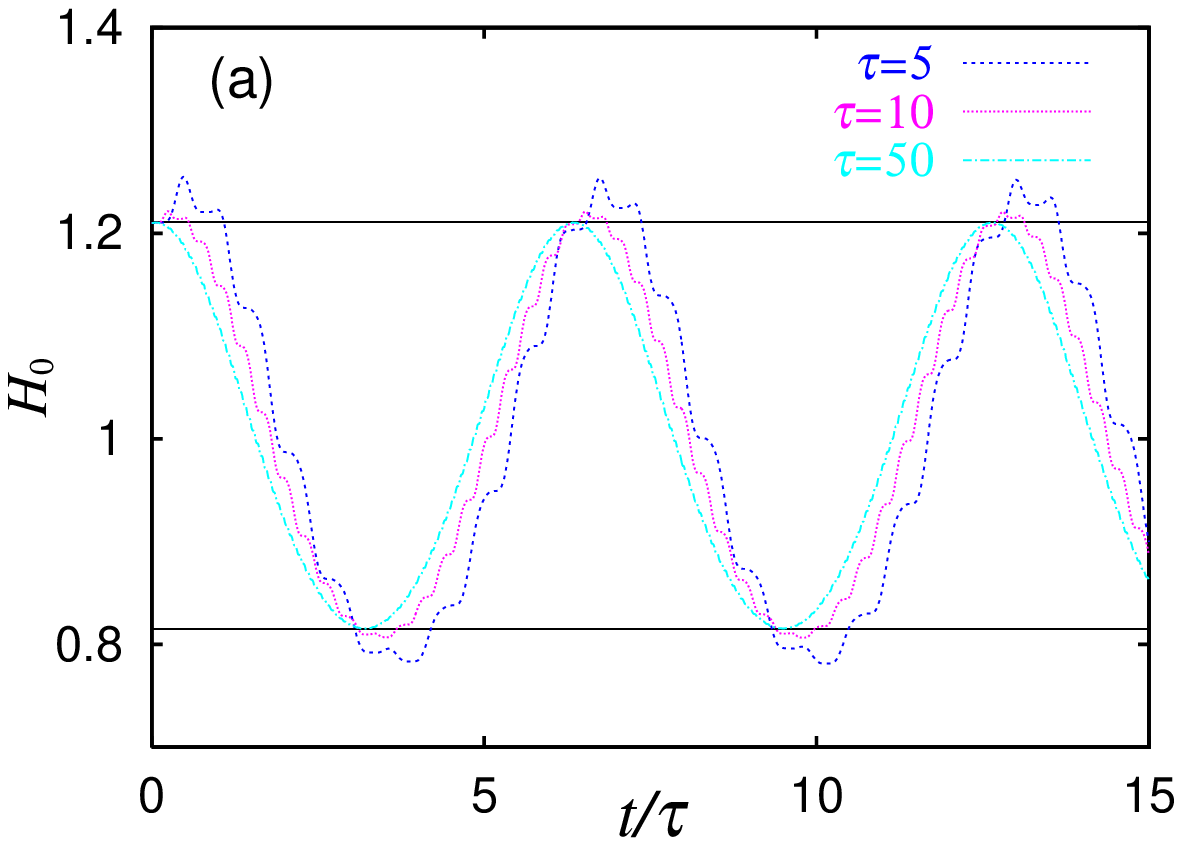}
\includegraphics[width=7cm]{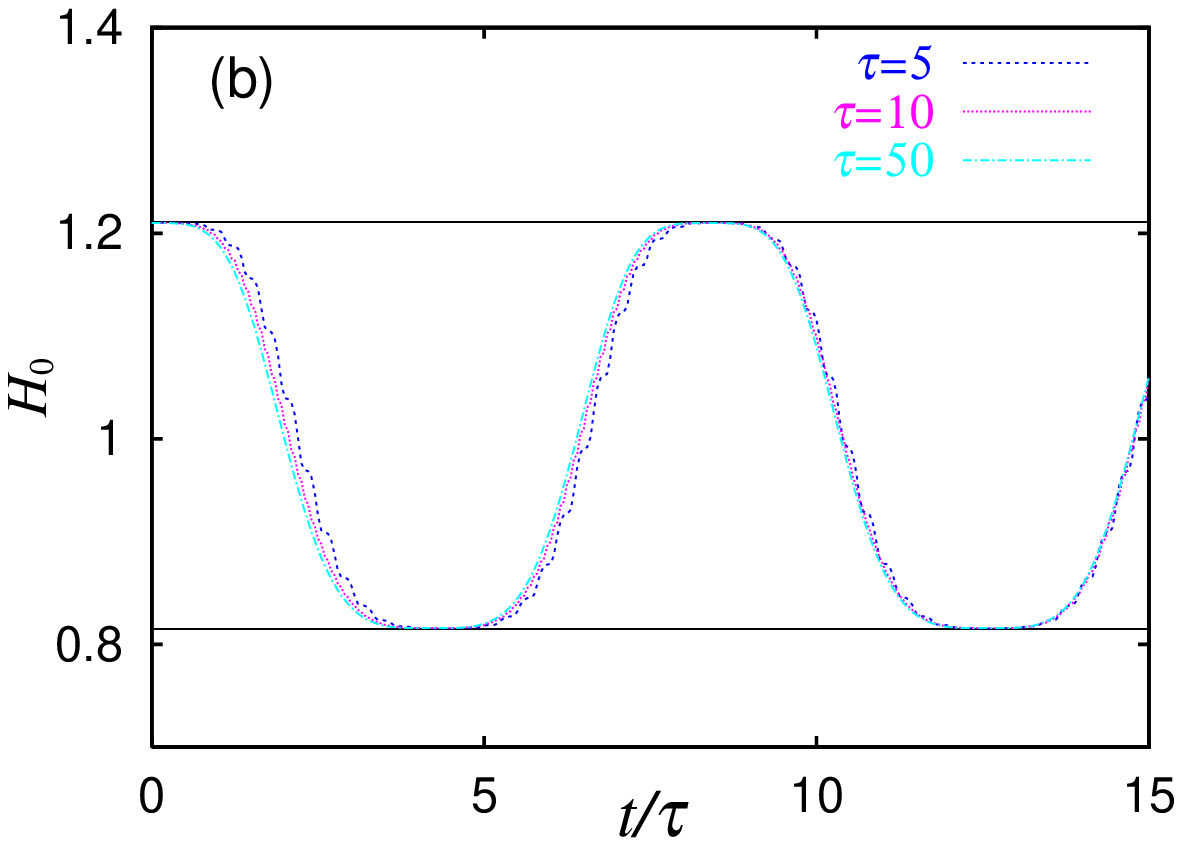}
\includegraphics[width=7cm]{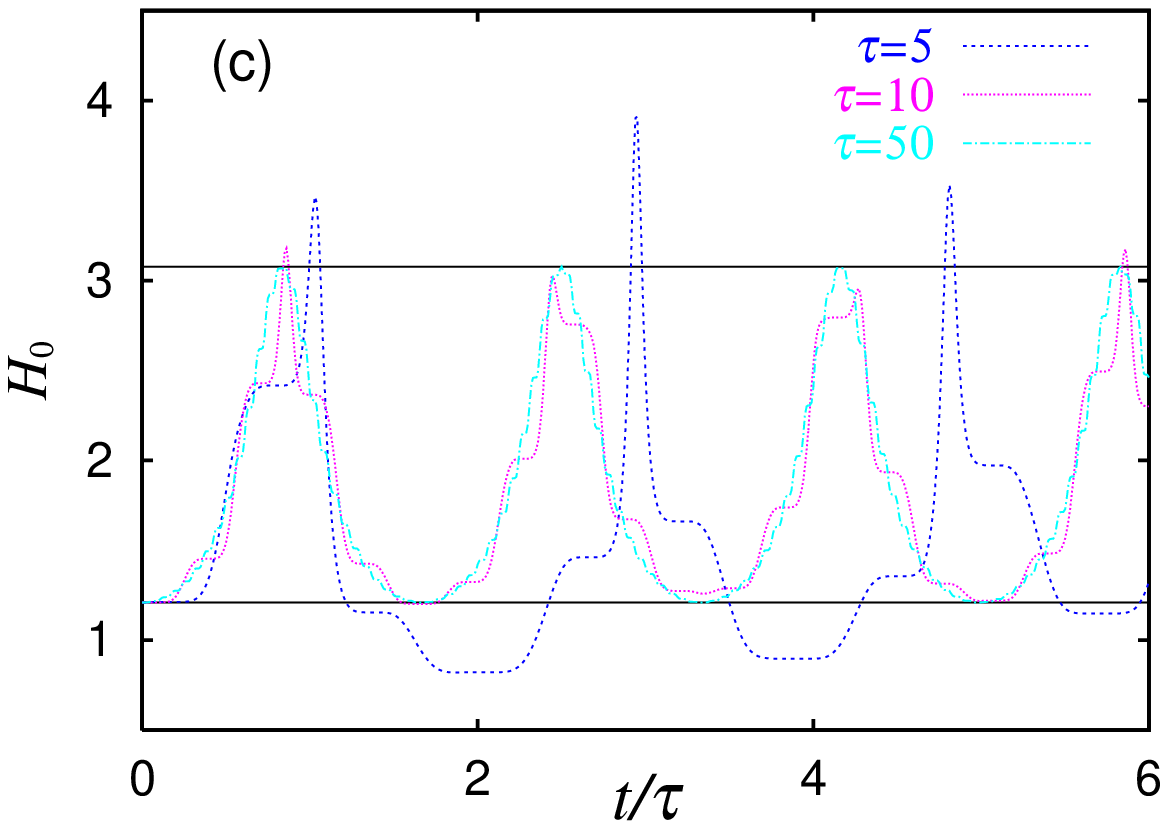}
\caption{
(Color online)~
Time evolutions of the energies of the systems with three thermostats
(a) $g = p\zeta$, (b) $g = p\zeta^3$, and (c) $g = p^3\zeta$.
Three cases $\tau = 5$, $10$, and $50$ are plotted in each figure.
The upper and the lower limits of energies
estimated from the inequality~(\ref{eq_inequality}) are shown as the solid lines.
The theoretical estimation becomes more accurate with larger $\tau$.
}
\label{fig_energy}
\end{figure}

In order to study the ergodicity of the present extended method,
we consider the one-dimensional harmonic oscillator described by the Hamiltonian $H_0 = p^2/2 + q^2/2$.
Then we have the following equations of motion:
\begin{eqnarray*}
\dot{p} &=& - q - p^{2m+1} \zeta^{2n+1}, \\
\dot{q} &=& p, \\
\dot{\zeta} &=& \displaystyle \frac{1}{\tau^2} z_n \left( p^{2m+2} - \frac{2m+1}{\beta}p^{2m} \right).
\end{eqnarray*}
Introducing the polar coordinates by $p = r \cos \theta$
and $q = r \sin \theta$,
we have the equations of motion in terms of $(r,\theta,\zeta)$ as
\begin{eqnarray*}
\dot{r} &=& - r^{2m+1} \zeta^{2n+1} \cos^{2m+2} \theta, \\
\dot{\theta} &=& 1 + r^{2m} \zeta^{2n+1} \cos^{2m+1} \theta \sin \theta, \\
\dot{\zeta} &=& \displaystyle \frac{1}{\tau^2} z_n  \left( r^{2m+2} \cos^{2m+2} \theta  - \frac{2m+1}{\beta}r^{2m} \cos^{2m} \theta \right).
\end{eqnarray*}
With large enough $\tau$, the variables $r$ and $\zeta$ vary much slower than $\theta$ does.
Therefore, we can replace $\cos^{2m} \theta$ with its average $c_m$ defined by
\begin{equation}
c_m \equiv \frac{1}{2\pi} \int_0^{2 \pi} \cos^{2m} \theta \diff \theta, \label{eq_cm}
\end{equation}
and we obtain the following two equations:
\begin{eqnarray*}
\dot{r} &=& - c_{m+1} r^{2m+1} \zeta^{2n+1}, \\
\dot{\zeta} &=& \displaystyle \frac{1}{\tau^2} z_n \left( c_{m+1} r^{2m+2}  -  \frac{2m+1}{\beta} c_mr^{2m}\right). 
\end{eqnarray*}
The above two equations lead to
\begin{equation}
- \tau^2 \frac{\zeta^{2n+1}}{z_n} \diff \zeta =
\left(
r - \frac{2m+1}{\beta} \frac{c_m}{c_{m+1}}\frac{1}{r} 
\right) \diff r.  \label{eq_separable}
\end{equation}
With a function $Z_n(\zeta)$ defined in
$$
\frac{\diff Z_n}{\diff \zeta} = \tau^2 \frac{\zeta^{2n+1}}{z_n}, 
$$
we have a conserved value
$H_0 - (m+1)\beta^{-1} \log H_0 + Z_n$
determined by the initial condition.
Here we have used the definition $H_0 \equiv r^2/2$
and the relation $(2m+1)c_m = 2(m+1) c_{m+1}$ obtained from the integration by parts of Eq.~(\ref{eq_cm}).
The function $Z_n(\zeta)$ has the minimum value $Z_n(0)$ at $\zeta = 0$
because $\diff Z_n/\diff \zeta <0$ if $\zeta <0$ and $\diff Z_n/\diff \zeta>0$ if $\zeta >0$.
Therefore, we have the following inequality:
\begin{equation}
H_0 - \frac{m+1}{\beta} \log H_0 \leq C \label{eq_inequality}
\end{equation}
with a constant $C$.
This inequality means that the energy of the system has the
minimum and the maximum values (see Fig.~\ref{fig_inequality}),
and that the system consequently loses its ergodicity.


In order to confirm our arguments, we study 
three thermostats, {\it i.e.}, $g = p\zeta$, $g = p\zeta^3$, and $g = p^3\zeta$.
All the thermostats are coupled with the one-dimensional harmonic oscillator
and the inverse temperature $\beta$ is set to be $1.0$.
For the relaxation time, we study three cases $\tau = 5$, $10$, and $50$.
The initial condition is set to be $(p,q,\zeta)=(1.1, 1.1, 0)$.
This condition gives $C = 1.019$ for $g = p\zeta$ and $g = p\zeta^3$,
and $C = 0.829$ for $g=p^3\zeta$.
From the inequality~(\ref{eq_inequality}),
we can estimate the range of the energy as follows:
\begin{eqnarray}
0.815 &\leq H_0& \leq 1.211 \quad (g = p\zeta, g = p\zeta^3), \label{eq_estimation_1} \\
1.210 &\leq H_0& \leq 3.076 \quad (g = p^3\zeta). \label{eq_estimation_2}
\end{eqnarray}
Time evolutions of the systems were numerically calculated by the fourth-order Runge--Kutta method
with the time step $0.005$ and those of the energies are shown in Fig.~\ref{fig_energy}.
The ranges of the energies agree well with our theoretical
estimation~(\ref{eq_estimation_1})~and~(\ref{eq_estimation_2}) for larger values of~$\tau$.


We have studied the ergodicity
of the thermostat family based on deterministic and time-reversible dynamics.
We have obtained the conserved value for the harmonic-oscillator system
coupled with the single-variable thermostats.
This conserved value causes the energy to be bounded,
and the system consequently loses its ergodicity.
We performed numerical simulations and have confirmed our theoretical arguments.
The conserved value exists in the system with the single additional variable,
since it is generally impossible to make a separable form as Eq.~(\ref{eq_separable})
for the system with two or more additional variables.
Therefore, the number of the additional degrees of freedom
is essential for the ergodicity of the system~\cite{WatanabeKobayashi}.
While we have studied the harmonic-oscillator system,
it is straightforward to apply our arguments for 
similar systems such as $H_0= p^2/2 + q^{2k}/2k~(k=1,2,3,\cdots)$.

We have given the general expression for the one-variable thermostats
with the pseudo friction term $g = p^{2m+1}\zeta^{2n+1}$.
For the case $n=0$, the thermostat can be regarded as a method controlling higher moments of the kinetic energy~\cite{KineticMoments}.
On the other hand, there are no clear physical interpretations for general cases $n \neq 0$.
Additionally, a more general form of pseudo Hamiltonian is available~\cite{Kusnezov1},
while we have assumed the quadratic form of the additional variable as in Eq.~(\ref{eq_pseudo_h}).
Therefore, it should be one of the further issues to clarify the physical meanings of 
general thermostats.


The authors would like to thank Miyashita, Ito, and Todo groups for fruitful discussions.
We also thank Sasa group for useful suggestions.
The present work was partially supported by the 21st COE program,
``Frontiers of Computational Science," Nagoya University.


\begin{thebibliography}{9}
\bibitem{Woodcock} L.\ V.\ Woodcock, {\it Chem. Phys. Lett.}, {\bf 10}, 257 (1971).
\bibitem{GaussianThermostat} W.\ G.\ Hoover, A.\ J.\ C.\ Ladd, and B.\ Moran, {\it Phys. Rev. Lett.}, {\bf 48}, 1818 (1982);
D.\ J.\ Evans, {\it J. Chem. Phys}, {\bf 78}, 3297 (1983).
\bibitem{Nose} S. Nos\'{e},  {\it Mol.\ Phys.}, {\bf 52}, 255 (1984).
\bibitem{NoseHoover}  W.\ G.\ Hoover, {\it Phys.\ Rev.\ A}, {\bf 31}, 1695 (1985).
\bibitem{NoseReview} S. Nos\'{e}, {\it Prog. Theor. Phys. Suppl.}, {\bf 103}, 1 (1991).
\bibitem{Dettmann} C.\ P.\ Dettmann and G.\ P.\ Morriss, {\it Phys.\ Rev.\ E}, {\bf 54}, 2495 (1996).
\bibitem{NosePoincare} S.\ D.\ Bond, B.\ J.\ Leimkuhler, and B.\ B.\ Lairdy, {\it J.\ Comput.\ Phys.}, {\bf 151}, 114 (1999).
\bibitem{HooverAoki} W.\ G.\ Hoover, K.\ Aoki, C.\ G.\ Hoover, and S.\ V.\ De Groot,
{\it Physica D}, {\bf 187}, 253 (2004).
\bibitem{Kusnezov2} D.\ Kusnezov, A.\ Bulgac, and W.\ Bauer, {\it Ann. Phys.} (N.~Y.), {\bf 214}, 180 (1992).
\bibitem{NoseHooverChain} G. J. Martyna, M. L. Klein, and M. Tuckerman, {\it J. Chem. Phys.}, {\bf 97}, 2635 (1992).
\bibitem{KineticMoments} W.\ G.\ Hoover and B.\ L.\ Holian, {\it Phys.\ Lett.\ A}, {\bf 211}, 253 (1996).
\bibitem{Kusnezov1} D.\ Kusnezov, A.\ Bulgac, and W.\ Bauer, {\it Ann. Phys.} (N.~Y.), {\bf 204}, 155 (1990).
\bibitem{Branka}
A.\ C.\ Bra\'nka and K.\ W.\ Wojciechowski, {\it Phys.\ Rev.\ E}, {\bf 62}, 3281 (2000);
A.\ C.\ Bra\'nka, M. Kowalik, and K.\ W.\ Wojciechowski, {\it J. Chem.\ Phys.}, {\bf 119}, 1929 (2003).
\bibitem{even} 
In the even number case, the simple expression for $\dot{\zeta}$ is obtained to be
$\dot{\zeta} = - \tau^{-2} \beta p ( p^2 - 2/\beta ) \zeta \exp{( \beta \zeta^2/2)}$
with $g(p,\zeta) = p^{2} \exp{(\beta \zeta^2/2)}$.
We performed a simulation with the above $g$, and found that the system was very unstable,
while the system became isothermal with appropriately chosen parameters.
\bibitem{WatanabeKobayashi} H.\ Watanabe and H.\ Kobayashi,
{\it Molecular Simulation}, {\bf 33}, 77 (2007).
\end{thebibliography}
\end{document}